\documentclass{article}
\usepackage{graphicx}
\usepackage{amsmath,amsthm,amsfonts}
\usepackage{numprint}
\npthousandsep{,}
\usepackage{xcolor}
\usepackage{booktabs}
\definecolor{citeblue}{rgb}{0.1,0,.4}
\usepackage{xurl}
\usepackage[%
,colorlinks=true%
,bookmarks=false%
,linkcolor=citeblue%
,citecolor=citeblue%
,plainpages=false]{hyperref}

\usepackage{listings}

\lstdefinelanguage{TPTP}{
  morekeywords={fof,cnf,include,axiom,conjecture,negated_conjecture},
  sensitive=true,
  morecomment=[l]{\%},      %
  morestring=[b]'           %
}

\author{Mikol\'a\v s Janota\\
Czech Technical University in Prague, CIIRC, Czechia}
\title{Experimental Results for Vampire on the Equational Theories Project}
\begin{document}
\maketitle
\begin{abstract}
	Equational Theories Project is a collaborative effort, which explores
	the validity of certain first-order logic implications of certain kind.
	The project has been completed but triggered further research. This
	report investigates how much can be automatically proven and disproven
	by the automated theorem prover Vampire. An interesting conclusion is
	that Vampire can prove all the considered implications that hold and
	also is able to refute a vast majority of those that do not hold.
\end{abstract}

\section{Introduction}
This report accompanies experiments carried out by the report
all the experiments and relevant script are placed on zendodo~\cite{zenodo}.

Terrence Tao proposed on his blog~\cite{blog} a collaborative project, which
aims to bring together mathematicians and researchers on automated reasoning and
other related fields. The project aims to classify all implications of a certain
type. The implications are between two universally quantified first order
logic  equalities that use a single binary operation---let us denote this
operation as~$*$. For illustration, consider the following implications.
\begin{align}
	(x*y)*z=x*(y*z) &\rightarrow  x*y=y*x\label{eq:1}\\
	x*y=y*x &\rightarrow  (x*y)*z=x*(y*z)\label{eq:2}\\
	x*y=u*w &\rightarrow  x*y=y*x\label{eq:3}
\end{align}

The first implication~\eqref{eq:1} asks if associativity implies commutativity.
This implication does not hold because for instance matrix multiplication is
associative but it is not commutative. One may also ask if
commutativity implies associativity~\eqref{eq:2}, which also does not hold
because for instance $\frac{x+y}{2}$ is
commutative but it is not associative. On the other hand,
implication~\eqref{eq:3} does hold, because the left-hand side requires that the
operation~$*$ always returns the same value, and therefore it is necessarily
commutative.

\section{Experimental Setup}
We consider all the equations in \texttt{generate\_eqs\_list.eqs}. There are
$n=\numprint{4694}$ equations, which means there are $n^2-n$ possible
implications pairs, giving~\numprint{22028942} pairs.
Even though many pairs could be  inferred from the value of other pairs through
the transitivity of implication, purposefully we do not do that.
Meaning, all pairs are targeted directly  and only marked as solved if the one
of the prover's configurations decided its validity (refuted/proven).

The automated theorem prover Vampire~\cite{vampire} supports saturation-based
proving~\cite{BachmairGanzinger1994,BachmairGanzinger2001}. It also has a
finite model builder~\cite{paradox,fmb},  which has the SAT solver CaDiCaL as
its backend~\cite{cadical}.\footnote{Enabled by the options
  \texttt{-sa fmb -sas cadical}.}

The input for Vampire was generated in the TPTP format~\cite{tptp},
in CNF,  already negated and skolemized.\footnote{The \texttt{generate\_tptp.py}
	script is used for this.} The following TPTP corresponds
to the test whether commutativity implies associativity.
\begin{lstlisting}
%
%
cnf(lhs, axiom, m(X, Y) = m(Y, X)).
cnf(rhs, negated_conjecture, m(a, m(b, c)) != m(m(a, b), c)).
\end{lstlisting}

All experiments were run on a server with two AMD EPYC 7513 32-Core processors @
3680 MHz and with 514\,GB\,RAM with 100 jobs in parallel. The problems were
tackled by Vampire 5.0.0 (Release build, commit 128f1f6ca on 2025-07-30 12:07:12
+0200) with CaDiCaL: cadical-2.1.3.
Times were measured  in walk-clock time.

\section{Experiments}
\begin{table}
  \caption{Solving methods}\label{tab:methods}
  \centering
  \begin{tabular}{ll}
    method &  command-line arguments \\\midrule
    fmb 500i &  \verb|-i 500 -sa fmb -sas cadical| \\
    satur 500i &  \verb|-i 500 --mode casc| \\
    fmb 60s &  \verb|-t 60s -sa fmb -sas cadical| \\
    satur 600s &  \verb|-t 600s --mode casc| \\
    fmb 600s &  \verb|-t 600s -sa fmb -sas cadical| \\
  \end{tabular}
\end{table}

\begin{table}[htbp]
  \centering
  \begin{tabular}[c]{lccc}
    \toprule
    Method & Refuted & Proven &   Total\\\midrule
    fmb 500i & \numprint{13837151} & \numprint{275209} & \numprint{14112360} \\
    satur 500i & \numprint{778} & \numprint{7895986} & \numprint{7896764} \\
    fmb 60s & \numprint{16302} & \numprint{0} & \numprint{16302} \\
    satur 600s & \numprint{36} & \numprint{2390} & \numprint{2426} \\
    fmb 600s & \numprint{28} & \numprint{0} & \numprint{28} \\
    \midrule total & \numprint{13854295} & \numprint{8173585} & \numprint{22027880} \\
    \bottomrule
  \end{tabular}
  \caption{Overview of the Results}\label{tab:results}\vspace{0pt}
\end{table}

\begin{figure}[h]
  \centering
  \includegraphics[width=.6\textwidth]{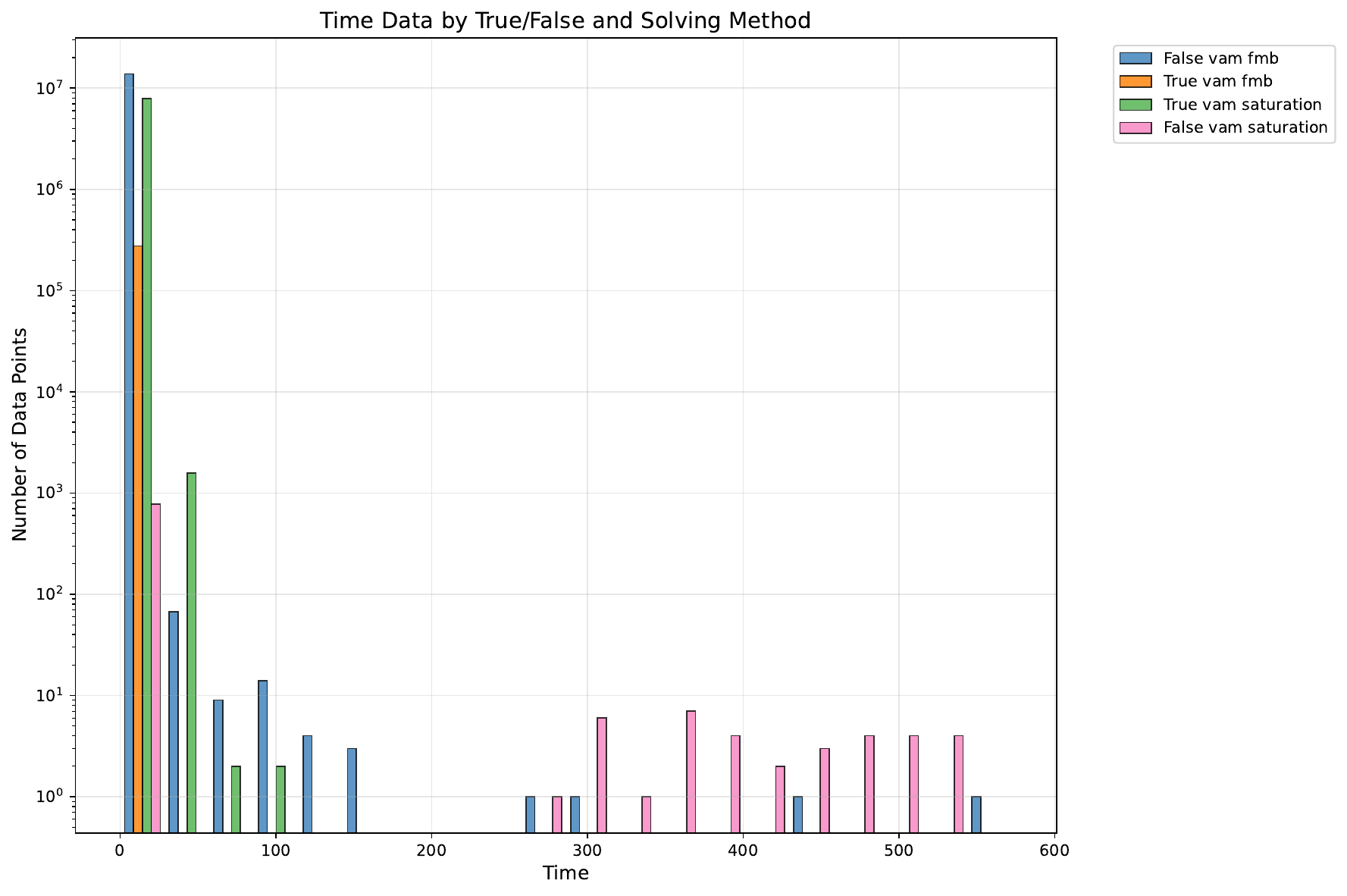}
  \caption{Histogram of the solving times.}\label{fig:runtimes}
\end{figure}

Vampire was run on all the problems with increasing time out and alternating
between the finite model building mode and the saturation mode as summarized by
Table~\ref{tab:methods}.

Table~\ref{tab:results} shows how many problems were solved by the individual
method. Most problems are solved with the short timeout of 500 instructions.
Figure~\ref{fig:runtimes} shows the solving times organized in a histogram and
divided by the solving method and the result.

As expected, the finite model builder is mainly successful in refuting
implications and the saturation-based approach in proving them. However, it can
also be the other way around even though it is rare.

When proving the implication $A\rightarrow B$, a saturation-based prover
\emph{can} determine that the implication does not hold. This is when the
prover runs on the formula $A\land\lnot B$ and eventually, the calculus of the prover
does not enable it to derive any more clauses, and it has not derived the empty
clause so far. In such case, however, we do not have a witnessing model that
would show that the implication does not hold (a model of $A\land\lnot B$). It
is also not guarantee that a finite model exists if this happens.

All the problems that were not solved are marked false in The equational
project~\cite{dashboard}. This means that Vampire can prove all the implications
that can be proven, and the challenge lies the implications that need to be
refuted. Additional 22 problems can be derived by calculating propagating via
transitivity of implication.\footnote{%
For example $511\not\rightarrow 3079$ follows from $1120\rightarrow 511$ and $1120\not\rightarrow 3079$.}
Only 310 of the undecided implications require an infinite model according to
the Equational project, which indicates there is also a room for improvement for
finite model finding. Infinite model finding of course poses a hard challenge.

\subsubsection*{Acknowledgements}
The author would like to thank Martin Suda, 
Geoff Sutcliffe, and Chad Brown for discussions about
this project.
The research was supported by the Ministry of Education, Youth and Sports
within the dedicated program ERC~CZ under the project \textsf{POSTMAN} no.~LL1902,
by the Czech Science Foundation grant no.~25-17929X, and
by the European Union under the project \textsf{ROBOPROX}
(reg.~no.~CZ.02.01.01/00/22\_008/0004590).
This article is part of the \textsf{RICAIP} project that has received funding from
the European Union's Horizon~2020 research and innovation programme under grant
agreement No~857306.

\bibliographystyle{plainurl}

\begin{thebibliography}{10}

\bibitem{dashboard}
Equational theories project.
\newblock URL: \url{https://teorth.github.io/equational_theories/dashboard/}.

\bibitem{BachmairGanzinger1994}
Leo Bachmair and Harald Ganzinger.
\newblock Rewrite-based equational theorem proving with selection and
  simplification.
\newblock {\em Journal of Logic and Computation}, 4(3):217--247, 1994.
\newblock Original formalization of the superposition calculus in a saturation
  framework.
\newblock \href {https://doi.org/10.1093/logcom/4.3.217}
  {\path{doi:10.1093/logcom/4.3.217}}.

\bibitem{BachmairGanzinger2001}
Leo Bachmair and Harald Ganzinger.
\newblock Resolution theorem proving.
\newblock In Alan Robinson and Andrei Voronkov, editors, {\em Handbook of
  Automated Reasoning}, volume~I, pages 19--99. Elsevier and {MIT} Press, 2001.
\newblock Foundational description of saturation-based reasoning and the
  superposition calculus.

\bibitem{cadical}
Armin Biere, Tobias Faller, Katalin Fazekas, Mathias Fleury, Nils Froleyks, and
  Florian Pollitt.
\newblock {CaDiCaL 2.0}.
\newblock In Arie Gurfinkel and Vijay Ganesh, editors, {\em Computer Aided
  Verification - 36th International Conference, {CAV} 2024, Montreal, QC,
  Canada, July 24-27, 2024, Proceedings, Part {I}}, volume 14681 of {\em
  Lecture Notes in Computer Science}, pages 133--152. Springer, 2024.
\newblock \href {https://doi.org/10.1007/978-3-031-65627-9\_7}
  {\path{doi:10.1007/978-3-031-65627-9\_7}}.

\bibitem{paradox}
Koen Claessen and Niklas S{\"o}rensson.
\newblock New techniques that improve {MACE}-style model finding.
\newblock In {\em Proc.\ of Workshop on Model Computation (MODEL)}, 2003.
\newblock URL:
  \url{https://web.archive.org/web/20070109072155/http://www.cs.chalmers.se/~koen/pubs/entry-model-paradox.html}.

\bibitem{vampire}
Laura Kov{\'a}cs and Andrei Voronkov.
\newblock First-order theorem proving and vampire.
\newblock In Natasha Sharygina and Helmut Veith, editors, {\em Computer Aided
  Verification}, pages 1--35, Berlin, Heidelberg, 2013. Springer Berlin
  Heidelberg.

\bibitem{fmb}
Giles Reger, Martin Riener, and Martin Suda.
\newblock Symmetry avoidance in mace-style finite model finding.
\newblock In Andreas Herzig and Andrei Popescu, editors, {\em Frontiers of
  Combining Systems - 12th International Symposium, FroCoS 2019, London, UK,
  September 4-6, 2019, Proceedings}, volume 11715 of {\em Lecture Notes in
  Computer Science}, pages 3--21. Springer, 2019.
\newblock \href {https://doi.org/10.1007/978-3-030-29007-8\_1}
  {\path{doi:10.1007/978-3-030-29007-8\_1}}.

\bibitem{zenodo}
Mikol\'a\v s~Janota.
\newblock Experimental results for vampire on the equational theories project,
  2025.
\newblock \href {https://doi.org/10.5281/zenodo.16910288}
  {\path{doi:10.5281/zenodo.16910288}}.

\bibitem{tptp}
Geoff Sutcliffe.
\newblock The {TPTP} problem library and associated infrastructure.
\newblock {\em J. Autom. Reasoning}, 43(4):337--362, 2009.
\newblock \href {https://doi.org/10.1007/s10817-009-9143-8}
  {\path{doi:10.1007/s10817-009-9143-8}}.

\bibitem{blog}
Terrence Tao.
\newblock A pilot project in universal algebra to explore new ways to
  collaborate and use machine assistance?, 2025.
\newblock URL:
  \url{https://terrytao.wordpress.com/2024/09/25/a-pilot-project-in-universal-algebra-to-explore-new-ways-to-collaborate-and-use-machine-assistance/}.

\end{thebibliography}

\end{document}